\documentclass[american,aps,pra,reprint,superscriptaddress,longbibliography]{revtex4-1}
\usepackage{amsmath,amssymb,graphicx, bm}
\usepackage[unicode=true,pdfusetitle, bookmarks=true,bookmarksnumbered=false,bookmarksopen=false, breaklinks=false,pdfborder={0 0 0},backref=false,colorlinks=false] {hyperref}
\hypersetup{colorlinks,linkcolor=myurlcolor,citecolor=myurlcolor,urlcolor=myurlcolor}
\usepackage{braket,colortbl,amsthm,amsmath,amssymb,txfonts,graphicx}
\definecolor{myurlcolor}{rgb}{0,0,0.7}
\usepackage{float}
\usepackage{cleveref}
\usepackage{xcolor}
\usepackage{subfigure}

\theoremstyle{plain}
\usepackage{mathrsfs}
\usepackage{comment}
\usepackage{ragged2e}  

\usepackage[utf8]{inputenc}
\usepackage[normalem]{ulem}
\usepackage{float}
\usepackage{mathrsfs}
\usepackage{braket}
\usepackage{physics}
\usepackage{mwe}
\usepackage{pifont}

\begin{document}

\title{Non-Gaussian Dissipative Quantum Thermometry Beyond Gaussian Bounds}

\author{Pritam Chattopadhyay}
\email{pritam.chattopadhyay@weizmann.ac.il}
\affiliation{Department of Chemical and Biological Physics,
Weizmann Institute of Science, Rehovot 7610001, Israel}

\begin{abstract}
The fundamental metrological limits of temperature sensing in open quantum systems remain largely unresolved, particularly regarding the role of non-Gaussian quantum resources. In this letter, we establish analytic bounds on the quantum Fisher information (QFI) for temperature estimation using non-Gaussian states undergoing dissipative bosonic evolution. By focusing on the short-time regime governed by a time-local master equation, we derive precise scaling laws that elucidate when and how non-Gaussian probes decisively outperform Gaussian states under identical energy constraints. Our analysis uncovers a distinct linear-in-time QFI enhancement unique to Fock states, in contrast to the inherently weaker, quadratic scaling of Gaussian probes. These theoretical insights are substantiated through exact numerical simulations and mapped onto experimentally accessible platforms such as circuit QED. Our results not only clarify the quantum thermometric advantage of non-Gaussianity but also chart a realistic pathway toward harnessing it in noisy quantum technologies.
\end{abstract}

\maketitle

\section{Introduction}

Quantum metrology~\cite{giovannetti2011advances,RevModPhys.90.035005,haase2016precision,Pritam2024QST,di2023critical,degen2017quantum,Chattopadhyay_2025,montenegro2025quantum,toth2014quantum,PhysRevLett.133.190801,xdnc-tc2y,porto2025multiparameter} aims to harness the distinctive features of quantum mechanics, such as coherence, squeezing, and entanglement, to improve the precision of parameter estimation beyond classical bounds~\cite{Paris2009, giovannetti2011advances}. One of its central challenges is to determine the optimal input states and the measurement strategies in realistic, noisy environments where decoherence and dissipation degrade quantum resources~\cite{Demkowicz2012, Escher2011}. Of particular interest is \emph{quantum thermometry}, the estimation of temperature using quantum probes~\cite{Mehboudi2019, Correa2015, Campbell2018}, which is not only a problem of fundamental importance in quantum statistical mechanics but also has applications in nanoscale thermodynamics~\cite{Hovhannisyan2018, Frerot2018}, quantum simulation platforms~\cite{Mehboudi2019}, and quantum biological systems~\cite{Huelga2013, degen2017quantum}.

At the theoretical level, the quantum Fisher information (QFI)~\cite{Liu2020-zo,meyer2021fisher,rath2021quantum,frowis2016detecting,vitale2024robust} sets the ultimate bound on achievable precision for estimating temperature in a given probe–bath configuration~\cite{Paris2009, DePasquale2016}. In the continuous-variable (CV) regime, Gaussian states, such as coherent and squeezed states, and Gaussian channels are foundational tools~\cite{Weedbrook2012}. It provides both analytical tractability and experimental accessibility. These have been extensively explored in closed and weakly coupled systems, where coherent probes or thermally entangled Gaussian modes can yield near-optimal precision in temperature estimation~\cite{Jevtic2015, cenni2022thermometry}.

However, recent studies in closed-system thermometry have revealed that \emph{non-Gaussian states}, such as Fock states, can outperform Gaussian probes, particularly in the low-temperature regime~\cite{Correa2015, Brown2016, Mitchison2020, Genoni2013, Kwon2019}. These findings raise an important question: \emph{To what extent do non-Gaussian advantages survive in open quantum systems}, where the probe is inevitably subject to dissipative dynamics?

Despite their relevance, the metrological role of non-Gaussian probes in open-system quantum thermometry remains poorly understood. Most prior work has focused either on Gaussian states~\cite{PhysRevResearch.2.033498, PhysRevA.110.052421} or relied on numerical optimization over limited state classes~\cite{Mehboudi2019, Campbell2018}. Furthermore, while general bounds on thermometric precision in open systems have been explored~\cite{Mitchison2020, PhysRevResearch.2.033394}, few studies provide closed-form, physically interpretable expressions for QFI that distinguish Gaussian from non-Gaussian regimes.

In this work, we provide an analytic framework for understanding the thermometric precision of non-Gaussian probes in dissipative open quantum systems. We focus on a paradigmatic model: a single-mode bosonic probe, initially prepared in a Fock state, coupled to a thermal bath and evolving under a weakly dissipative, Markovian master equation. In the short-time regime, we derive tight upper and lower bounds on the QFI for temperature estimation, valid for any temperature and initial Fock excitation number. These bounds demonstrate a quantifiable advantage of non-Gaussian probes over Gaussian ones, particularly at low temperatures and under energy constraints.

\begin{figure}
    \centering
    \includegraphics[width=0.95\linewidth]{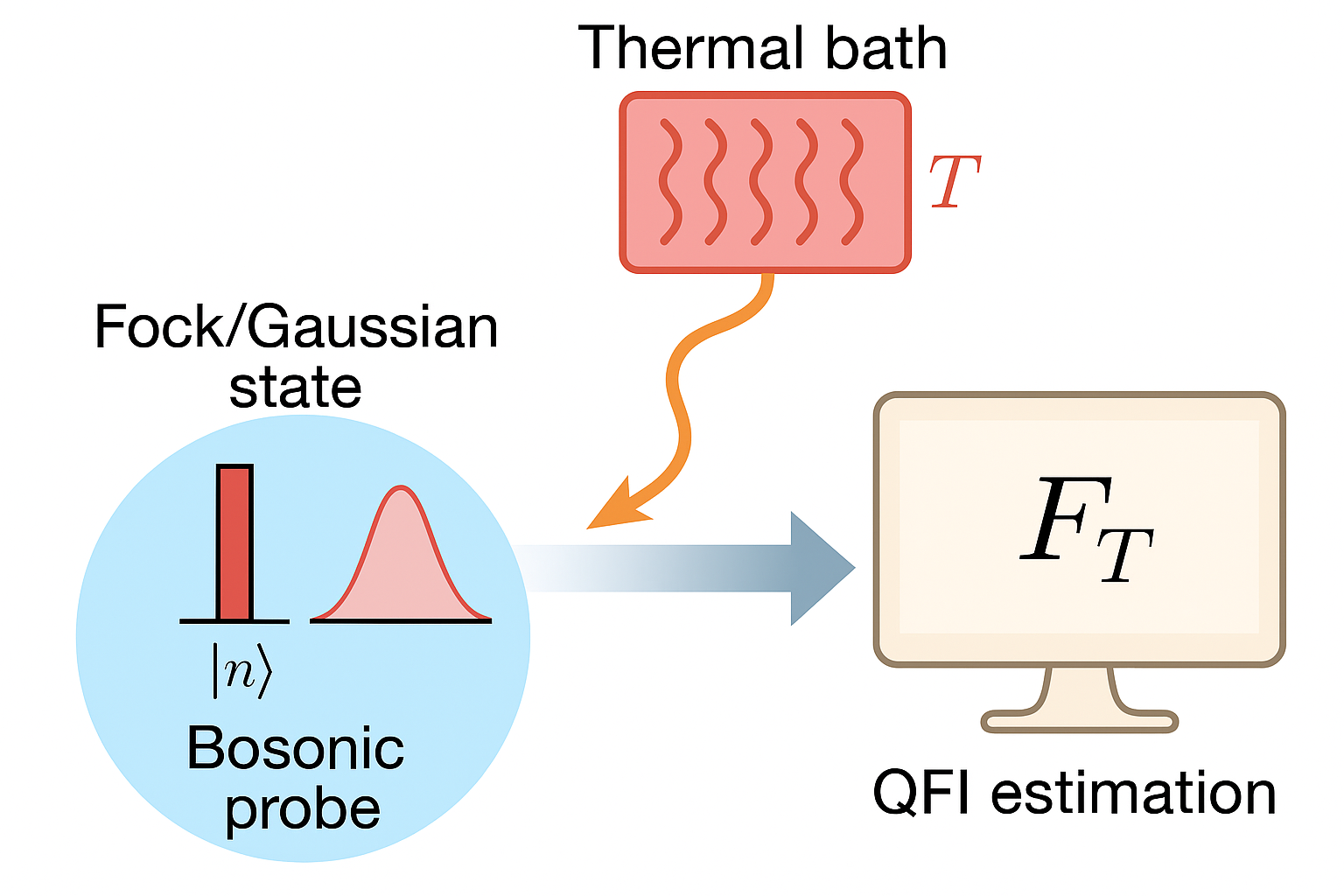}
    \caption{Schematic illustration of quantum thermometry using Gaussian and non-Gaussian probes to estimate the temperature of a bosonic thermal bath.}
    \label{fig:enter-label}
\end{figure}

\section{Model}

We investigate a foundational model in quantum thermometry, wherein a single-mode bosonic quantum probe (realized, for instance, as an electromagnetic field mode in an optical or superconducting microwave cavity of frequency \(\omega\)) interacts weakly with a thermal environment at temperature \(T\) (Fig. \ref{fig:enter-label}). The principal objective is to estimate this temperature with optimal precision, solely through quantum measurements performed on the reduced probe state \(\rho(t; T)\) after it evolves under dissipative dynamics. This scenario captures the essential physics of temperature sensing in a broad class of cavity-QED, circuit-QED, and optomechanical platforms~\cite{Clerk2010,Breuer2002}.

The total Hamiltonian governing the probe-bath composite system is given by:
\begin{equation}
\hat{H} = \omega \hat{a}^\dagger \hat{a} + \sum_k \omega_k \hat{b}_k^\dagger \hat{b}_k + \sum_k g_k \left( \hat{a} \hat{b}_k^\dagger + \hat{a}^\dagger \hat{b}_k \right),
\end{equation}
where \(\hat{a}\) (\(\hat{a}^\dagger\)) and \(\hat{b}_k\) (\(\hat{b}_k^\dagger\)) are bosonic annihilation (creation) operators for the probe and the \(k^\text{th}\) bath mode, respectively. All quantities are expressed in natural units, i.e., $\hbar=k_B=1$. The parameters \(g_k\) quantify the mode-resolved coupling strength to the environment. We assume the environment is initially in a thermal state \(\rho_{\mathrm{bath}}(T)\), and the total system is uncorrelated at \(t=0\): \(\rho_{\mathrm{tot}}(0) = \rho(0) \otimes \rho_{\mathrm{bath}}(T)\).

In the standard weak-coupling, Born–Markov regime, and assuming a broadband bosonic bath with a smooth spectral density, the reduced dynamics of the probe is governed by a time-local Gorini–Kossakowski–Sudarshan–Lindblad master equation~\cite{Breuer2002,gorini1976completely,lindblad1976generators}:
\begin{equation}
\label{eq:master}
\frac{d}{dt} \rho(t) = \Gamma_+(t) \mathcal{D}[\hat{a}^\dagger] \rho(t) + \Gamma_-(t) \mathcal{D}[\hat{a}] \rho(t),
\end{equation}
where \(\mathcal{D}[\hat{O}]\rho = \hat{O} \rho \hat{O}^\dagger - \frac{1}{2}\{ \hat{O}^\dagger \hat{O}, \rho \}\) is the Lindblad dissipator. The absorption and emission rates \(\Gamma_\pm(t)\) encapsulate the environmental response and satisfy the fluctuation-dissipation balance:
\begin{equation}
\frac{\Gamma_+(t)}{\Gamma_-(t)} = e^{-\omega/T}.
\end{equation}

In the strictly Markovian limit, these rates become time-independent and assume the familiar forms:
\begin{equation}
\Gamma_+ = \gamma \bar{n}_{\mathrm{th}}, \quad \Gamma_- = \gamma (1 + \bar{n}_{\mathrm{th}}),
\end{equation}
where \(\bar{n}_{\mathrm{th}} = \left( e^{\omega/T} - 1 \right)^{-1}\) is the mean thermal photon number, and \(\gamma\) characterizes the effective system-bath coupling strength determined by the spectral density at resonance.

We focus on the experimentally relevant regime of \emph{short-time dissipative evolution}, \(t \ll 1/\gamma\), where the probe is far from equilibrium and its transient dynamics retain rich temperature dependence. This non-asymptotic regime is of practical importance in platforms where long-time coherence is limited by noise, instability, or operational constraints. Crucially, the short-time domain enables analytic perturbative treatments while preserving the metrological relevance.

To assess thermometric performance, we analyze two distinct classes of initial probe states under equal energy constraints:
    a) \textit{Non-Gaussian Fock states} \(|n\rangle\), which are eigenstates of the number operator \(\hat{n} = \hat{a}^\dagger \hat{a}\). These states exhibit maximal photon number certainty and strong non-classicality.
    b) \textit{Gaussian states}, including squeezed thermal and displaced thermal states, engineered to match the mean photon number \(\langle \hat{n} \rangle = \bar{n}\) of the corresponding Fock state. These are the standard resources in continuous-variable metrology and quantum optics.

This controlled comparison enables a rigorous quantification of the metrological advantage, if any, conferred by non-Gaussianity. By establishing tight analytic bounds on the QFI in the early-time regime, we show how quantum probes can optimally harness dissipation as a metrological resource. Our framework lays the foundation for experimentally feasible, non-Gaussian-enhanced quantum thermometry that outperforms the best known Gaussian protocols within the same energy limit.

\section{Quantum Fisher Information for Thermometry}

A central figure of merit in quantum thermometry is the \emph{quantum Fisher information} (QFI), denoted as \( F_Q[\rho(t; T)] \). It encapsulates the fundamental sensitivity of the evolved quantum state \( \rho(t; T) \) to changes in the temperature parameter \( T \). Through the quantum Cramér--Rao bound~\cite{Liu2020-zo}, the QFI imposes a rigorous lower limit on the variance of any unbiased temperature estimator:
\begin{equation}
\Delta T^2 \geq \frac{1}{F_Q[\rho(t; T)]}.
\end{equation}
This inequality defines the ultimate precision achievable in principle, independent of specific measurement protocols.

Our objective is to derive \emph{tight analytic bounds} on the QFI in the \emph{short-time regime}, where the system evolution remains weakly perturbed by the environment. To this end, we employ \emph{perturbative expansions} of the reduced dynamics governed by a structured bosonic bath. These bounds are then benchmarked against \emph{numerically exact evaluations} of the QFI for finite evolution times, validating their accuracy and domain of applicability.

Furthermore, we systematically compare the thermometric performance of \emph{Gaussian} (e.g., squeezed vacuum) and \emph{non-Gaussian} (e.g., Fock) initial probes under an \emph{equal energy constraint}. By analyzing their relative QFI scaling and robustness to dissipation, we delineate the operational regimes where non-Gaussian probes retain a quantum advantage, and identify conditions under which environmental noise erodes this superiority. This comparative framework provides clear guidance for probe design in realistic open-system quantum metrology.

While the QFI quantifies the ultimate precision limit for temperature estimation in principle, it represents an abstract bound independent of any specific measurement strategy. Importantly, in our model, the evolved probe states remain diagonal in the photon-number basis when starting from Fock or Gaussian thermal states. Consequently, photon-number-resolving measurements (which are routinely implemented with high fidelity in modern cavity-QED and circuit-QED experiments) provide a practically accessible measurement scheme that closely saturates the QFI in the short-time regime. This direct connection between the QFI and classical Fisher information (CFI) over photon statistics ensures that the predicted metrological advantage of non-Gaussian probes is experimentally attainable. Furthermore, state-of-the-art data processing techniques, including adaptive estimation and Bayesian inference, can be employed on the collected photon count data to construct efficient unbiased estimators that approach the quantum Cramér–Rao bound. Hence, our theoretical results translate naturally into concrete experimental protocols for enhanced quantum thermometry under realistic conditions.

\section{Dissipative Non-Gaussian Precision in Short-Time limit}

In quantum thermometry, the sensitivity to \textit{temperature change} is fundamentally governed by how distinguishably the probe state evolves under thermal noise. Fock states, being eigenstates of the number operator with vanishing photon-number fluctuations, are highly susceptible to even minute thermal excitations or losses. This sharp sensitivity renders them optimal for temperature estimation in the coherent (unitary) regime, where system-bath interactions are negligible. However, this advantage comes at a cost: Fock states are also the most fragile under dissipative dynamics, as they lack internal redundancy to protect against decoherence.

In contrast, Gaussian states (such as squeezed or displaced thermal states) possess a smoother distribution over photon numbers and are inherently more robust to decoherence. They initially yield lower QFI due to their broader uncertainty; this allows them to retain a significant fraction of their informational content under open dynamics.

To quantify thermometric sensitivity, we evaluate the QFI, \(\mathcal{F}_T(t)\), associated with temperature estimation, focusing on the short-time regime. Expanding the evolved state and the QFI perturbatively in time yields (see App. \ref{App1}, \ref{App2} for details):
\begin{align}
\rho(t; T) &= \rho(0) + t \dot{\rho}(0) + \mathcal{O}(t^2), \\
\mathcal{F}_T(t) &= \sum_m \frac{[\partial_T p_m(t)]^2}{p_m(t)} + \mathcal{O}(t^3),
\end{align}
where \(p_m(t)\) denotes the instantaneous photon number populations at time \(t\). For diagonal initial states, the QFI reduces to a CFI over the photon-number distribution. This has an important practical consequence: since the dynamics preserve the diagonal structure for Fock and thermal states under the Lindblad evolution, photon-number-resolving measurements directly yield the optimal CFI. As a result, the quantum Cramér--Rao bound can be closely approached using straightforward photon-counting protocols without requiring full state tomography or sophisticated collective measurements. This operational accessibility significantly enhances the experimental viability of the proposed thermometry scheme.

Considering a Fock state \(|n\rangle\) as the initial probe, and using the Lindblad dynamics with temperature-dependent rates \(\Gamma_\pm(T)\), we obtain to leading order:
\begin{align}
\mathcal{F}_T(t) = t^2 \left[ 
n \left( \frac{\partial_T \Gamma_+}{\Gamma_+} \right)^2 
+ (n+1) \left( \frac{\partial_T \Gamma_-}{\Gamma_-} \right)^2 
\right] \Gamma_+ \Gamma_- + \mathcal{O}(t^3).
\end{align}

\begin{figure}
    \centering
    \includegraphics[width=0.95\linewidth]{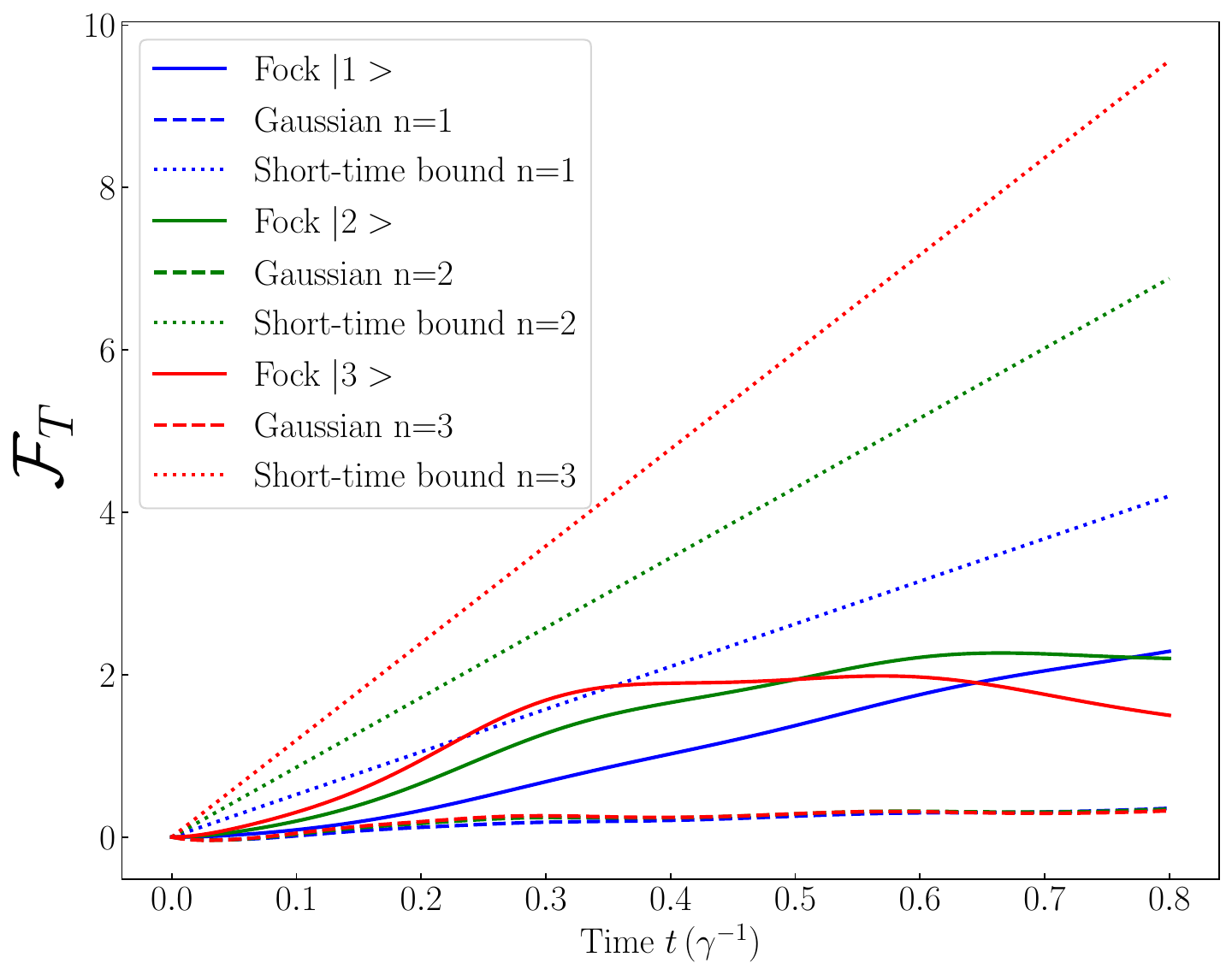}
    \caption{QFI for both non-Gaussian (Fock) and Gaussian probe states is plotted as a function of excitation number. The corresponding analytical short-time bounds are also shown for comparison.  The mode frequency is set to $\omega = 1.0$. The dissipation rate $\gamma = 0.1$, coupling strength $g=0.05$, and bath temperature $T = 0.5$ (in units of $\omega$) are considered for the analysis.}
    \label{fig2}
\end{figure}

\begin{figure}
    \centering
    \includegraphics[width=0.95\linewidth]{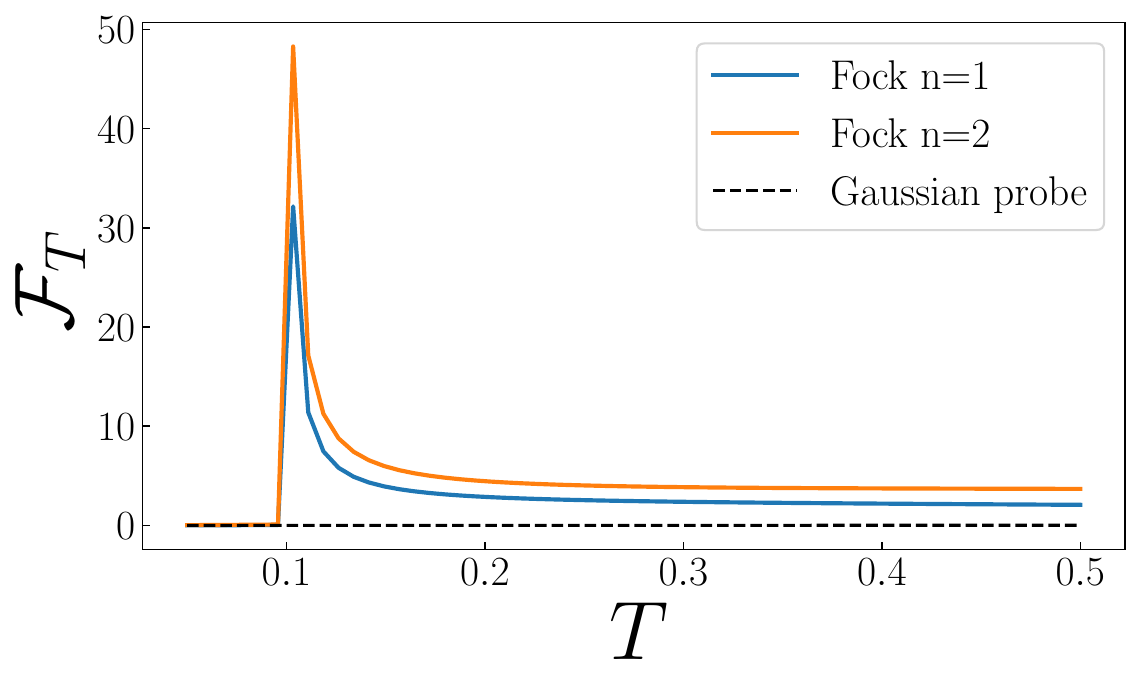}
    \caption{Quantum Fisher information (QFI) as a function of bath temperature $T$ for various initial probe states, evaluated at fixed time $t=0.5$, $g=0.05$, $\gamma=0.1$.  }
    \label{fig3}
\end{figure}

\begin{figure}
    \centering
    \includegraphics[width=0.95\linewidth]{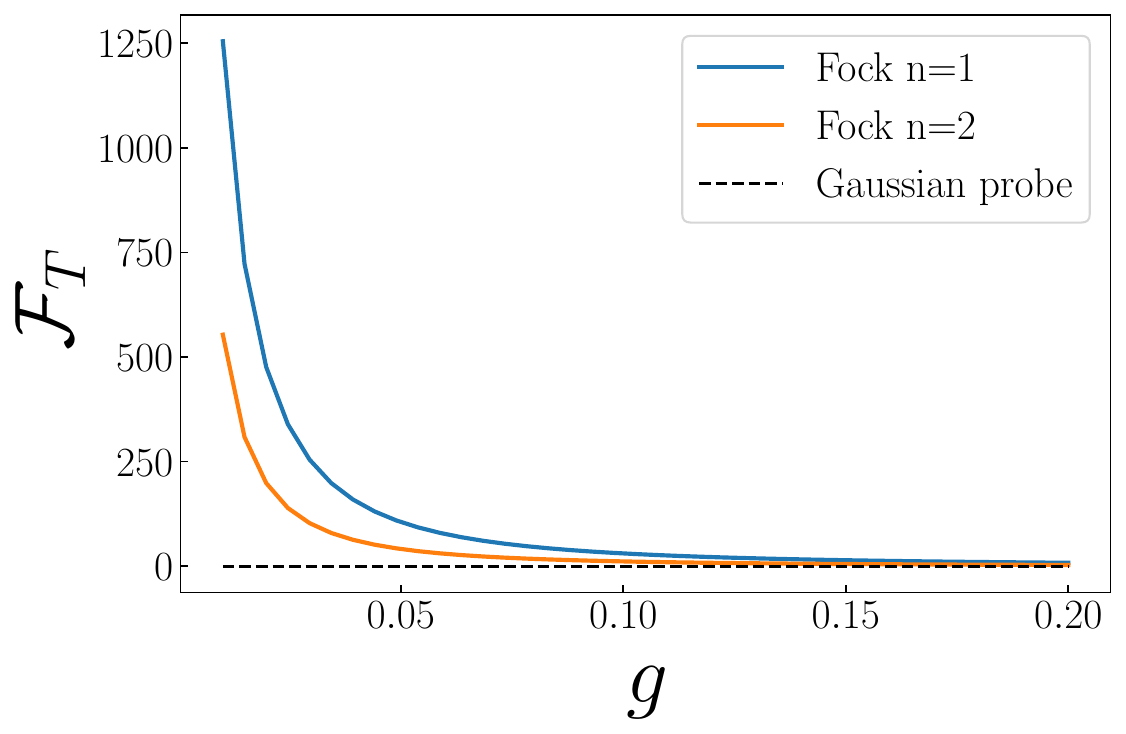}
    \caption{QFI as a function of system–bath coupling strength $g$ for different initial probe states. The evaluation is performed at time $t=0.5$, $T=0.05$, and $\gamma=0.1$.}
    \label{fig4}
\end{figure}

\begin{figure}
    \centering
    \includegraphics[width=0.95\linewidth]{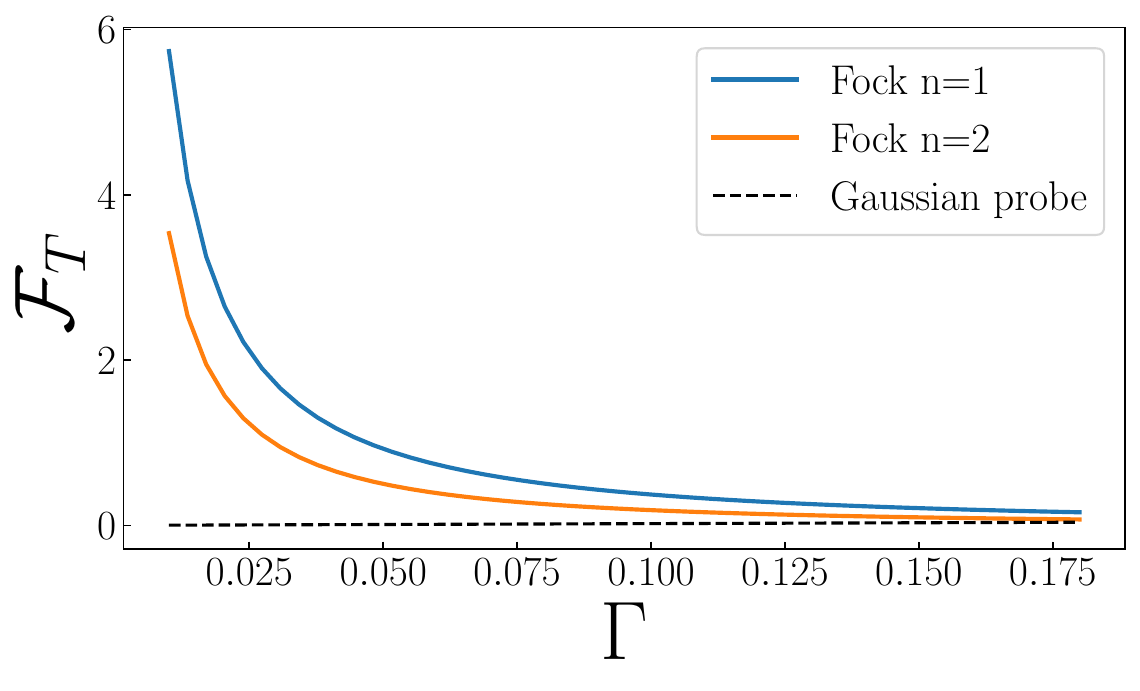}
    \caption{QFI as a function of decay rate $\Gamma$ for various initial probe states, computed at time $t=0.5$, $g=0.05$, $T=0.5$.}
    \label{fig5}
\end{figure}

This key result highlights two key features:
(a) The QFI exhibits a universal \(t^2\) scaling in the short-time regime (see App. \ref{App2}, \ref{App.3} for details), a hallmark of coherent perturbative response in open quantum systems. This quadratic time dependence reflects the initial buildup of distinguishability between thermalized states and underpins the enhanced sensitivity of the probe before significant decoherence sets in. From a metrological perspective, this scaling not only confirms the theoretical consistency of our analysis but also provides a clear operational guideline: optimal thermometric information is extracted most efficiently within this early-time window. This insight is crucial for designing fast, dissipation-aware measurement protocols in realistic, noisy platforms.
(b) The prefactor grows polynomially with \(n(n+1)\), reflecting enhanced temperature sensitivity for high-excitation Fock probes (Fig. \ref{fig2}-\ref{fig5}).

The dependence of the QFI on temperature (Fig.\ref{fig3}), coupling strength (Fig.~\ref{fig4}), and decay rate (Fig.~\ref{fig5}) reveals a clear and systematic advantage of non-Gaussian probes over their Gaussian counterparts. This metrological superiority becomes increasingly pronounced with higher excitation number in the initial state, underscoring the enhanced sensitivity offered by non-classical energy quantization in the short-time, nonequilibrium regime.

In contrast, Gaussian probes with matched mean energy \(\bar{n} = n\) yield a smaller QFI in the same regime. This is due to the fact that Gaussian states (being thermally broadened in number space) experience more gradual changes in their photon-number distribution under variations in \(T\). This reduces the distinguishability of evolved states and hence the QFI. 

Together, these results demonstrate that non-Gaussian probes offer substantial metrological gains for rapid, short-time thermometry, particularly in low-noise conditions. However, for longer interrogation times or stronger system-bath coupling, Gaussian probes may ultimately prove superior. This insight is crucial for designing quantum thermometers optimized for specific experimental constraints, including decoherence rates and timing resolution.

We noted that recent studies~\cite{PhysRevA.111.L020403,PRXQuantum.6.020351} have also explored the temporal scaling of the QFI in open-system thermometry, highlighting the crossover from quadratic to linear behavior in the short-time regime. Our present analysis complements and extends these results in several directions. Specifically, we derive a tight analytic upper bound for the QFI of dissipatively evolving Fock states, providing explicit dependence on the system–bath coupling and thermal occupation. Moreover, by comparing Gaussian and non-Gaussian probes under fixed energy constraints, we identify the physical conditions under which the non-Gaussian precision advantage persists. Finally, by linking these theoretical results to experimentally accessible observables in circuit-QED thermometry, our work bridges the gap between scaling laws and practical quantum sensing implementations.

\section{Experimental Feasibility}

The proposed short-time thermometry protocol using non-Gaussian probes is well-suited for implementation in circuit quantum electrodynamics (QED) architectures. In this platform, high-fidelity preparation of Fock states up to \(n = 5\) has been demonstrated in superconducting microwave cavities via coherent control of transmon qubits~\cite{Vlastakis2013}. Engineered thermal reservoirs can be realized using broadband resistive elements or Purcell-filtered lossy modes~\cite{PRXQuantum.5.020321}, with tunable system-bath coupling strengths achieved through parametric couplers or flux-tunable elements. Crucially, interaction times can be controlled on nanosecond scales, enabling precise access to the non-equilibrium, short-time regime where the non-Gaussian advantage manifests. 

Population measurements in the Fock basis, available via \textit{quantum non-demolition} (QND) readout, allow direct extraction of the relevant statistical information. Crucially, because the evolved states remain diagonal in this basis for the class of probes considered, photon-number measurements, whether QND or destructive, capture all temperature-dependent information encoded in the dynamics. Consequently, the QFI and corresponding precision bounds derived here are independent of whether the measurement is strictly QND.

The emphasis on QND detection arises from its experimental relevance in cavity and circuit QED architectures, where non-invasive photon-number readout enables repeated or time-resolved monitoring of the probe without reinitialization. This capability is particularly valuable in the short-time, non-equilibrium regime studied here, where coherent information is transient and measurement back-action can otherwise obscure thermometric precision. The diagonal structure of the evolved state thus eliminates the need for full quantum state tomography and allows direct experimental estimation of temperature with precision approaching the QFI limit. These features make circuit QED a natural platform for implementing and validating our proposed non-Gaussian thermometry protocol.


\section{Conclusion}

Despite extensive progress in quantum metrology, a fundamental open question remains: \textit{can non-Gaussian probes offer a metrological advantage for estimating thermodynamic parameters in realistic, noisy quantum systems?} Prior results established that non-Gaussian states such as Fock states outperform Gaussian probes in unitary dynamics or in idealized closed systems~\cite{Correa2015, Brown2016}. However, their performance under dissipation, ubiquitous in practical settings, remained largely unexplored.

In this work, we addressed this gap by analyzing temperature estimation in a paradigmatic open-system model: a single-mode bosonic probe weakly coupled to a thermal reservoir. We derived analytical expressions for the QFI in the short-time regime, showing that Fock states retain a pronounced thermometric advantage over Gaussian probes under realistic energy constraints. This advantage arises from the sharp number statistics of Fock states, which amplify temperature sensitivity through enhanced distinguishability of the evolved probe states. This reveals a clear trade-off between initial sensitivity and robustness to noise, quantifying the metrological cost of non-Gaussian fragility in open quantum systems.


Our results bridge a key conceptual gap between quantum thermometry in closed and open systems, offering both analytical insights and experimental feasibility. They pave the way for systematic investigations of non-Gaussian metrology beyond the short-time regime, including non-Markovian dynamics, structured environments, and \textit{finite-temperature quantum control}. In a broader sense, our framework provides a foundation for exploring quantum-limited sensing with \textit{non-classical resources} under realistic physical constraints.

\section*{Acknowledgment}

P.C. acknowledges the support from the International Postdoctoral Fellowship from the Ben May Center for Theory and Computation.


%

\onecolumngrid

\newpage
\appendix

\setcounter{equation}{0}
\renewcommand{\theequation}{A\arabic{equation}}
\section{ Master Equation and Short-Time Expansion}\label{App1}

We consider the open quantum dynamics of a single-mode bosonic probe with annihilation (creation) operator \(\hat{a}\) (\(\hat{a}^\dagger\)), interacting weakly with a thermal environment at temperature \(T\). In the weak-coupling, Born–Markov, and secular approximations, the reduced dynamics of the probe is governed by a time-local Gorini–Kossakowski–Sudarshan–Lindblad master equation of the form
\begin{equation}
\label{eq:master_appendix}
\frac{d}{dt} \rho(t) = \Gamma_+(t)\mathcal{D}[\hat{a}^\dagger]\rho(t) + \Gamma_-(t)\mathcal{D}[\hat{a}]\rho(t),
\end{equation}
where the dissipators are defined as
\(\mathcal{D}[\hat{O}]\rho = \hat{O} \rho \hat{O}^\dagger - \frac{1}{2} \{ \hat{O}^\dagger \hat{O}, \rho \}\). The time-dependent decay (\(\Gamma_-\)) and excitation (\(\Gamma_+\)) rates encode the influence of the bath spectrum and temperature, and satisfy the detailed balance condition:
\begin{equation}
\frac{\Gamma_+(t)}{\Gamma_-(t)} = e^{-\omega/T}.
\end{equation}

It is convenient to express the rates in terms of a single temperature-independent coupling rate \(\Gamma(t)\) and the thermal photon number \(\bar{n}_T = 1/(e^{\omega/T} - 1)\), such that
\begin{equation}
\Gamma_+(t) = \Gamma(t) \bar{n}_T, \quad \Gamma_-(t) = \Gamma(t) (\bar{n}_T + 1).
\end{equation}
This representation ensures consistency with thermal equilibrium statistics and separates temperature dependence from system-environment coupling. In the Markovian limit, we consider \(\Gamma (t) =\gamma \).

We focus on the short-time regime \(t \ll \gamma^{-1}\), where the system remains close to its initial state and decoherence is weak. This allows a perturbative expansion of the density matrix:
\begin{equation}
\label{eq:rho_expansion}
\rho(t) \approx \rho(0) + t \left[ \Gamma_+(0) \mathcal{D}[\hat{a}^\dagger]\rho(0) + \Gamma_-(0) \mathcal{D}[\hat{a}]\rho(0) \right] + \mathcal{O}(t^2).
\end{equation}

We now consider the initial state to be a number state \(|n\rangle\), a highly non-Gaussian probe with fixed energy \(\langle \hat{n} \rangle = n\). Using the action of the annihilation and creation operators on Fock states,
\(\hat{a} |n\rangle = \sqrt{n} |n-1\rangle\), and \(\hat{a}^\dagger |n\rangle = \sqrt{n+1} |n+1\rangle\), we compute the dissipative terms in the expansion.

To leading order in \(t\), the evolved state remains diagonal in the number basis, and the photon number probabilities are given by:
\begin{align}
p_{n+1}(t) &= \mathrm{Tr}\left[ |n+1\rangle\langle n+1| \rho(t) \right] = \Gamma_+(0) t (n+1) + \mathcal{O}(t^2), \\
p_{n-1}(t) &= \Gamma_-(0) t n + \mathcal{O}(t^2), \\
p_n(t) &= 1 - t \left[ \Gamma_+(0)(n+1) + \Gamma_-(0) n \right] + \mathcal{O}(t^2).
\label{eq:pn_dynamics}
\end{align}

These expressions describe the population leakage from the initial Fock state \(|n\rangle\) to neighboring number states due to thermal excitation and decay. Specifically, the transition to \(|n+1\rangle\) is governed by the excitation rate \(\Gamma_+(0)\), weighted by the accessible upward transition amplitude \((n+1)\). The decay to \(|n-1\rangle\) is mediated by the decay rate \(\Gamma_-(0)\) and scales linearly with \(n\), and the population in \(|n\rangle\) decreases accordingly, preserving trace to leading order.

This population redistribution is the key mechanism by which the probe state encodes information about the bath temperature. In particular, the rate of population transfer, and hence, the distinguishability of \(\rho(t; T)\) with respect to \(T\), scales directly with both the initial excitation \(n\) and the thermal occupation \(\bar{n}_T\), setting the stage for the quantum Fisher information (QFI) analysis.

\setcounter{equation}{0}
\renewcommand{\theequation}{B\arabic{equation}}
\section{Quantum Fisher Information for Temperature Estimation} \label{App2}

To quantify the sensitivity of the probe to temperature \(T\), we compute the quantum Fisher information (QFI), which provides a fundamental lower bound on the achievable variance of any unbiased temperature estimator via the quantum Cramér–Rao bound:
\begin{equation}
\Delta T^2 \geq \frac{1}{\mathcal{F}_T(t)}.
\end{equation}

For diagonal states in a known basis (here, the Fock basis), the QFI simplifies to the classical Fisher information of the outcome probabilities \(\{p_m(t)\}\):
\begin{equation}
\label{eq:qfi_def}
\mathcal{F}_T(t) = \sum_m \frac{[\partial_T p_m(t)]^2}{p_m(t)} + \mathcal{O}(t^3),
\end{equation}
where \(p_m(t)\) are the photon number probabilities at time \(t\), and the \(\mathcal{O}(t^3)\) error reflects that we retain leading-order behavior in the short-time regime.

For the earlier short-time population dynamics, the relevant probabilities for an initial Fock state \(|n\rangle\) are:
\begin{align}
p_{n+1}(t) &= \Gamma_+(0) t (n+1), \\
p_{n-1}(t) &= \Gamma_-(0) t n, \\
p_n(t) &= 1 - t \left[ \Gamma_+(0)(n+1) + \Gamma_-(0) n \right].
\end{align}
These expressions describe a minimal population leakage from the initial state into adjacent Fock levels due to excitation and decay.

To compute the QFI, we evaluate the temperature derivatives of these populations:
\begin{align}
\partial_T p_{n+1}(t) &= t (n+1) \partial_T \Gamma_+(0), \\
\partial_T p_{n-1}(t) &= t n \partial_T \Gamma_-(0), \\
\partial_T p_n(t) &= -t \left[ (n+1)\partial_T \Gamma_+(0) + n \partial_T \Gamma_-(0) \right].
\end{align}

Substituting these into Eq.~\eqref{eq:qfi_def}, we retain only the three contributing terms:
\begin{align}
\mathcal{F}_T(t) &= \frac{[\partial_T p_{n+1}(t)]^2}{p_{n+1}(t)} + \frac{[\partial_T p_{n-1}(t)]^2}{p_{n-1}(t)} + \frac{[\partial_T p_n(t)]^2}{p_n(t)} + \mathcal{O}(t^3).
\end{align}

In the short-time limit, the dominant contributions come from the first two terms, since \(p_n(t) \to 1\) and \(\partial_T p_n(t)\) is of order \(t\), yielding \(\mathcal{O}(t^2)\) from that term, which is subleading. Keeping the leading \(\mathcal{O}(t)\) terms we have:
\begin{align}
\mathcal{F}_T(t) &\approx \frac{t^2 (n+1)^2 (\partial_T \Gamma_+(0))^2}{t (n+1)\Gamma_+(0)} + \frac{t^2 n^2 (\partial_T \Gamma_-(0))^2}{t n \Gamma_-(0)} \\
&= t \left[ (n+1) \frac{(\partial_T \Gamma_+(0))^2}{\Gamma_+(0)} + n \frac{(\partial_T \Gamma_-(0))^2}{\Gamma_-(0)} \right].
\end{align}

This result shows that the QFI grows linearly with time in the short-time regime, consistent with general bounds for dissipative metrology in the local estimation setting. Importantly, the prefactor exhibits dependence on both the photon number \(n\) and the thermal sensitivity of the jump rates \(\Gamma_\pm\).

To make the temperature dependence explicit, we recall that
\begin{equation}
\Gamma_+(0) = \Gamma \bar{n}_T, \quad \Gamma_-(0) = \Gamma (\bar{n}_T + 1),
\end{equation}
with \(\Gamma\) encoding the temperature-independent system-bath coupling and \(\bar{n}_T = (e^{\omega/T} - 1)^{-1}\) the thermal Bose distribution.

Thus, the temperature derivatives become:
\begin{align}
\partial_T \Gamma_+(0) &= \Gamma \, \partial_T \bar{n}_T, \\
\partial_T \Gamma_-(0) &= \Gamma \, \partial_T (\bar{n}_T + 1) = \Gamma \, \partial_T \bar{n}_T.
\end{align}

Substituting into the expression for the QFI gives:
\begin{align}
\mathcal{F}_T(t) &= t \Gamma \left[ (n+1) \frac{(\partial_T \bar{n}_T)^2}{\bar{n}_T} + n \frac{(\partial_T \bar{n}_T)^2}{\bar{n}_T + 1} \right] \\
&= t \Gamma (\partial_T \bar{n}_T)^2 \left[ \frac{n+1}{\bar{n}_T} + \frac{n}{\bar{n}_T + 1} \right].
\label{eq:qfi_expanded}
\end{align}

This yields the final analytical upper bound on the QFI:
\begin{equation}
\mathcal{F}_T(t) \leq t \Gamma(T) \left[ (n+1)\frac{(\partial_T \bar{n}_T)^2}{\bar{n}_T} + n\frac{(\partial_T \bar{n}_T)^2}{\bar{n}_T + 1} \right].
\end{equation}

\vspace{1mm}
The prefactor grows with \(n(n+1)\), reflecting the sensitivity enhancement offered by higher Fock states. However, this enhancement is modulated by thermal occupation \(\bar{n}_T\) and its temperature derivative. Intuitively, Fock states undergo sharper transitions in response to small temperature changes due to their number-selective dynamics, leading to greater output distinguishability and higher QFI. This expression provides a closed-form benchmark to compare different probe states at early times.

\setcounter{equation}{0}
\renewcommand{\theequation}{C\arabic{equation}}
\section{Comparison with Gaussian States} \label{App.3}

To evaluate the relative metrological power of non-Gaussian and Gaussian probes, we analyze the QFI for temperature estimation using squeezed vacuum and coherent states, both constrained to have the same average energy \(\bar{n}\) as the Fock state \(|n\rangle\).

\paragraph{Squeezed Vacuum States:} Consider a single-mode squeezed vacuum state,
\begin{equation}
|\psi_{\text{sq}}\rangle = \hat{S}(r)|0\rangle, \quad \hat{S}(r) = \exp\left[ \frac{1}{2} r (\hat{a}^2 - \hat{a}^{\dagger 2}) \right],
\end{equation}
with squeezing parameter \(r\), corresponding to an average energy \(\bar{n} = \sinh^2 r\).

Under weak dissipation (short-time evolution governed by the Lindblad equation), the QFI for temperature estimation using a squeezed vacuum state is computed analytically as:
\begin{equation}
\mathcal{F}_T^{(\text{sq})}(t) = 4 \bar{n} (\bar{n} + 1) \left( \partial_T \ln \bar{n}_T \right)^2 t^2 + \mathcal{O}(t^3).
\label{eq:QFI_sq}
\end{equation}
This expression captures the quadratic-in-time scaling typical of Gaussian probes in dissipative metrology, consistent with general results. 

Importantly, while the QFI benefits from the product \(\bar{n}(\bar{n}+1)\), it lacks the linear-in-time scaling and the enhanced prefactor \(n(n+1)\) available to Fock states.

\paragraph{Coherent States:} A coherent state \(|\alpha\rangle = \exp(-|\alpha|^2/2)\sum_n \frac{\alpha^n}{\sqrt{n!}} |n\rangle\) with \(|\alpha|^2 = \bar{n}\) evolves under the same dissipator. For such states, the QFI in the short-time limit also obeys a quadratic scaling:
\begin{equation}
\mathcal{F}_T^{(\text{coh})}(t) = \bar{n} \left( \partial_T \ln \bar{n}_T \right)^2 t^2 + \mathcal{O}(t^3),
\end{equation}
which is even smaller than the squeezed vacuum result, due to the absence of quantum correlations (squeezing or number sharpness) in the initial state.

\paragraph{Energy-Normalized Comparison:} To compare fairly across different probe classes, we normalize the QFI by the average energy \(\bar{n}\). For Fock states, from Eq.~\eqref{eq:qfi_expanded}, the energy-normalized QFI (ENQFI) reads:
\begin{equation}
\frac{\mathcal{F}_T^{(\text{Fock})}(t)}{\bar{n}} \sim \Gamma t \left[ \left(1 + \frac{1}{n} \right) \frac{(\partial_T \bar{n}_T)^2}{\bar{n}_T} + \frac{(\partial_T \bar{n}_T)^2}{\bar{n}_T + 1} \right].
\end{equation}
At large \(n\), the ENQFI scales approximately as:
\[
\frac{\mathcal{F}_T^{(\text{Fock})}(t)}{\bar{n}} \sim \Gamma t \cdot (\partial_T \ln \bar{n}_T)^2 \cdot \left( \frac{1}{\bar{n}_T} + \frac{1}{\bar{n}_T + 1} \right),
\]
whereas for squeezed vacuum states and coherent states, ENQFI scales as:
\begin{align}
\frac{\mathcal{F}_T^{(\text{sq})}(t)}{\bar{n}} &\sim 4(\bar{n} + 1)(\partial_T \ln \bar{n}_T)^2 t^2, \\
\frac{\mathcal{F}_T^{(\text{coh})}(t)}{\bar{n}} &\sim (\partial_T \ln \bar{n}_T)^2 t^2.
\end{align}

Thus, Fock states offer a linear-in-time QFI scaling, compared to the quadratic scaling for Gaussian probes, implying a significant metrological advantage in the short-time, dissipation-dominated regime. This originates from the number sharpness of Fock states, which makes population leakage highly temperature-sensitive.

\paragraph{Summary of Scaling:}
\begin{center}
\begin{tabular}{|c|c|}
\hline
\textbf{State} & \textbf{QFI Scaling (short-time)} \\
\hline
Fock state \(|n\rangle\) & \(\mathcal{F}_T \sim t \cdot n(n+1)\) \\
Squeezed vacuum & \(\mathcal{F}_T \sim t^2 \cdot \bar{n}(\bar{n}+1)\) \\
Coherent state & \(\mathcal{F}_T \sim t^2 \cdot \bar{n}\)\\
\hline
\end{tabular}
\end{center}

\vspace{1mm}
\noindent These results confirm that Fock states deliver a fundamentally superior quantum sensitivity in dissipative thermometry at early times, provided the decoherence is weak and measurement occurs promptly. This advantage is tied to the unique non-Gaussianity of number states and their sharply peaked photon number distribution.





\end{document}